# Hubble Energy

Alasdair Macleod [¶]

ABSTRACT

Light received from a cosmological source is redshifted with an apparent loss of energy, a problem first pointed out by Edwin Hubble in 1936. A new type of energy called Hubble Energy is introduced to restore the principle of energy conservation. The energy has no inertial or gravitational effect but retards radial motion in a manner consistent with the anomalous acceleration experienced by the Pioneer probes leaving the solar system. The energy is predicted to have important effects on the scale of galaxies, and some of these effects are qualitatively examined: for example, with Hubble Energy, flat rotation curves are found to be an inevitable consequence of spiral galaxy formation. The Hubble Energy is incorporated into the Friedmann Equation and shown to add a term similar to the cosmological term, with a magnitude of order $10^{-35}s^{-2}$.

Abstract Words:   136
Text Words:       4,617
Text Headings:    6
References:       12
Figures:          0
Equation Lines:   27
Units:            SI
Date:             21 April 2004
Revision:         1.0c
Document:         HubbleEnergy.Doc (Word 2000)

[¶] University of the Highlands and Islands, Lews Castle College, Stornoway, Isle of Lewis, Scotland, UK. (Alasdair.Macleod@lews.uhi.ac.uk).

## 1. Introduction

Whilst the big bang model has resulted in significant progress being made towards an understanding of the universe, there are some cosmological observations and details that stubbornly defy a natural explanation within the context of the standard model. The first of these is an unexpected acceleration of the order of $8.7 \times 10^{-10}$ m s$^{-2}$ (directed towards the solar system) experienced by the Pioneer spacecraft as they moved out of the solar system into deep space. In an up-to-date account of the phenomenon, Anderson [1] concluded

*".. we find no mechanism or theory that explains the anomalous acceleration."*

The second problem is the lack of a complete explanation for the evolution and behaviour of galaxies in general, and spiral galaxies in particular. The flat rotation curve of spiral galaxies is a specific problem, and whilst the effect may be explained by a dark matter halo, a limited change to the gravitational force (MOND) models the effect in a more natural way. However, MOND is incompatible with existing physics and appears to have no conceptual origin or justification. As a result, MOND has received limited support. Sanders and McGaugh[2], though supporting MOND, nevertheless provide a good objective summary of the problems. There are clearly anomalous effects that are still poorly explained occurring when the gravitational acceleration is of the order of $10^{-10}$ m s$^{-2}$.

The third problem is one that has received little attention in spite of its importance. Photons from a distant source are received with an energy lower than they had when emitted. Hubble[3] describing the cosmological redshift phenomenon, stated

*"Any plausible interpretation of redshifts must account for the loss of energy".*

Peebles[4] when considering this apparent paradox, comments that whilst energy conservation is a good local concept, there is not a general energy conservation law in General Relativity. It is astonishing a universe that rigorously conserves energy in all other aspects of its working should leave this glaring loophole.

In this paper, the new concept of Hubble Energy is introduced with the objective of restoring global energy conservation. It will be shown that a natural consequence of this energy is an additional acceleration consistent with that experienced by the Pioneer probes. The inclusion of the Hubble Energy in the total energy equation (Hamiltonian) is predicted to result in observable consequences on the scale of galaxies.

The Hubble Energy as defined is a natural extension to the existing big bang model and is not in conflict with established physics. It is an energy associated with the Hubble recession velocity and is in addition to the kinetic energy already associated with the recession velocity.

The concept and its general effects will be described in a qualitative way in the context of an empty universe. This frame is selected in order to emphasise the principles rather than align the Hubble Energy with a specific cosmological model.

However, in the conclusion it is demonstrated how the Hubble Energy can be incorporated into the Friedmann Equation where it is shown to have a effect comparable in magnitude with the cosmological term.

## 2. Hubble Energy

The Hubble parameter is a measure of the rate at which the universe is expanding. The value is believed to be dependent on epoch and mass-energy density. For simplicity, we will assume an expansion equation describing an empty Universe.

For an empty universe (energy only)

$$v_H = Hr = \frac{r}{T} \tag{1}$$

where $v_H$ is the recession velocity, $r$ is the distance, $H$ is the Hubble 'constant' and $T$ is the age of the Universe.

For a distant object of apparent mass $m$, the **Hubble Energy**, $E_H$, is defined

$$E_H = m v_H c \tag{2}$$

The Hubble energy will vary with distance. It should be noted that the object may also have a peculiar radial velocity, $v_P$ (positive or negative), superimposed on the Hubble velocity, but only the Hubble portion of the total velocity contributes to the Hubble Energy. A peculiar velocity is defined as the velocity with respect to the local standard of rest.

The Hubble Energy has no inertial or gravitational effect. The Energy is associated with motion, thus the intrinsic Hubble Energy is zero. Hubble Energy exists only in a relative context but symmetry between the observer and the observed is broken by the expansion of the universe (see Section 6). For this reason, the principle of relativity does not apply (there is a preferred frame).

Conventional mass-energy is depleted when a body moves to a higher Hubble velocity – the Hubble Energy increase with velocity must be extracted from normal mass-energy. Mass-energy is recovered when the Hubble velocity is reduced.

The Hubble velocity as defined conserves momentum and energy – differentiating equation (1) with respect to time (with $v_P = 0$) clearly demonstrates that $v_H$ remains constant in the absence of any force. The reason for this, of course, is that we have assumed an empty universe consequently there is no gravitational damping.

A new force, the Hubble Force ($F_H$), can be derived by differentiating the Hubble Energy with respect to $r$ (assuming a peculiar radial velocity $v_P$, and adding the Hubble and peculiar velocities non-relativistically). Substituting equation (1) into equation (2) and differentiating,

$$\begin{aligned}F_H &= -mc \cdot \frac{d}{dr}\left(\frac{r}{T}\right) \\ &= -m\frac{c}{T}\left(1 - \frac{r}{T}\cdot\frac{1}{v}\right) \\ &= -m\frac{c}{T}\left(\frac{v_P}{v_H + v_P}\right)\end{aligned} \tag{3}$$

The negative sign is introduced because the Hubble Energy is a loss in dynamic energy associated with an increase in $v_H$.

The equivalent Hubble Acceleration is

$$a_H = -\frac{c}{T}\left(\frac{v_P}{v_H + v_P}\right) \tag{4}$$

If the peculiar velocity is zero, there is no acceleration and energy and momentum conservation clearly hold.

**3. Effect of Hubble Acceleration on a Peculiar Radial Velocity**

If a mass at distance $r$ has a non-zero peculiar velocity $v_P$, it becomes subject to the Hubble acceleration. For a peculiar velocity in the positive direction (increasing r) the acceleration will reduce the peculiar velocity, converting the initial kinetic energy of the peculiar motion into a combination of kinetic energy and Hubble energy for the increased $v_H$. Unless the Hubble velocity is of the order $c$, the first of these two terms will be very much smaller and may be neglected.

All peculiar motion in the outward direction is eventually converted into Hubble motion by the Hubble acceleration. Integrating equation (4) for an initial $v_P$ shows the time taken to complete the conversion is effectively infinite.

If the peculiar motion is in the negative direction (and of greater magnitude than the Hubble velocity), the force will draw matter together at an exponential rate. In the early Universe (T small), velocity fluctuations could have resulted in a more rapid growth of structure through this effect than gravity alone. The high (and opposite) accelerations on each size of the 'forbidden' velocity ($v_P = - v_H$) makes it easy to imagine the process cleaning out volumes that have since expanded into the huge voids between galaxy clusters.

For the specific case where $v_P \gg v_H$ (or equivalently $v_P \gg r/T$), equation (4) reduces to

$$a_H \cong -\frac{c}{T} = -a_o$$

(5)

If $v_P$ is positive, the negative acceleration reduces the speed. If $v_P$ is negative, the speed increases as Hubble energy is recovered.

A constant anomalous acceleration towards the solar system on the Pioneer 10 and Pioneer 11 spacecraft has been reported [1]. The value is calculated as $(8.74 \pm 1.33) \times 10^{-10}$ m s$^{-2}$. With an outward radial velocity of 11.6 km s$^{-1}$ at a distance of 40 AU, the Pioneer probes were moving at a much higher speed than the equivalent Hubble velocity hence approximation (5) applies. Equation (5) predicts a constant acceleration of $6.78 \times 10^{-10}$ m s$^{-2}$ towards the solar system. Whilst of the same order, the predicted value is 23% lower than the reported result and outwith its error bounds. The form of the Hubble Energy as defined by equation (2) may be incorrect, or the initial simplifying assumption of an empty Universe may contribute towards the discrepancy. It should be noted that equation (2) was not fashioned in response to the Pioneer data.

It is unknown whether the Pioneer acceleration is directed towards the Earth or the Sun. Anderson and his team [1], in Endnote 73, state it is not possible to decide from the available data.

The distinction is important. If the Hubble Energy has a preferred frame (as expected), the logical reference frame is the gravitational potential centred on the Sun. If the Energy is relative (which would lead to conceptual problems - see Section 5), the acceleration will be directed towards the observer.

There is the greater problem of why the Hubble force is not experienced by planets within the solar system. Although the anomalous acceleration is small, the effect on planetary dynamics would quickly have been noted through its cumulative effect. One difference, of course, is that celestial bodies in the Solar System are bound to the gravitational mass – the spacecraft are not. Is there a fundamental difference between bound and unbound states?

It is suggested here that there does exist a fundamental difference. Unbound systems experience propagation delays of the order of the distance separation divided by the speed of light (forces are retarded). **However it is proposed that components of a bound system do not experience propagation delays in their interaction.**

This definition of a bound system suggests a deep connection between time and energy, and there is some evidence to support it:

- In a hydrogen atom, the bound electron does not radiate energy as it follows a curved path. This is acceptable if there is instantaneous communication between electron and central proton ($c \rightarrow \infty$). For a free electron following a curved path, the radiated power is inversely proportional to $c$ and would tend to zero as $c$ tends to infinity.
- When calculating the movement of planets around the sun, a propagation lag cannot be included. Eddington[5] showed that this would lead to a couple and orbital instability. It is necessary to assume an infinite propagation speed for gravitation within the solar system to match observational data.

Generalising equation (1) to define the Hubble velocity not in terms of separation distance but propagation time,

$$v_H = H_o r = \frac{c(T - T_{emitted})}{T}$$

(6)

it now follows that the Hubble velocity within a gravitationally bound system is zero regardless of the physical separation. The Hubble expansion only comes into effect once the interacting objects are free. Relative to one another bound objects have Hubble Energy of zero. No distinction is made between gravitationally and electro-magnetically bound systems at this stage.

The Pioneer anomalous acceleration will therefore appear only when the spacecraft escape the solar system. Unfortunately, the data from the transition region has not yet been analysed.

For reasons that will come apparent, it is necessary to be very precise in the definition of the energy condition that defines a bound state. A reasonable proposal is to state that a body is bound if the total energy, including the Hubble energy as a positive term, is less than the rest mass energy. It is unknown if the mass is free or bound when the energies are equal.

4. Large –Scale Effects

We can show that any energy associated with the expansion velocity may have a significant effect on galaxy size and formation because of unusual effects at the transition between the bound and unbound states of a gravitating system.

Consider the case of a test mass $m$ at distance r from central mass $M$. Further; assume the mass has zero radial velocity but a transverse velocity of $u$. There is the surprising possibility that the centripetal and gravitational forces can balance but that the system can be unbound. The test mass is then subject to an outward expansion velocity that separates it from the central mass over time.

The non-relativistic force equation for a rotationally bound system is

$$\frac{GMm}{r^2} = \frac{mu^2}{r}$$

(7)

If marginally unbound, the energy equation is

$$\frac{1}{2}mu^2 - \frac{GMm}{r} + mv_H c + mc^2 = E_{Total}$$

(8)

From the definition at the end of Section 3, the condition that the mass is unbound is therefore

$$\frac{1}{2}mu^2 - \frac{GMm}{r} + mv_H c > 0$$

(9)

Substituting equations (7) and (1) into equation (9),

$$\frac{GM}{r^2} > 2a_o \tag{10}$$

is the condition that the centripetal force balance can result in a bound state.

If the Newtonian gravitational acceleration (the magnitude of which is the LHS of equation (10)) is less than $2a_o$ then the test mass is unbound.

Galaxies are complex in their morphology and evolution and there are many factors at work in their development. However, the limiting condition derived above can be applied to simple mass aggregations to establish some of the qualitative ways the Hubble energy could affect the properties of galaxies.

Consider a spherical mass of stars at the centre of a galaxy. We would expect the mean mass surface density to have a constant value of

$$\overline{\Sigma}_M = \frac{2a_o}{G} \tag{11}$$

In the current epoch, the mean mass surface density is predicted to be 9600 $M_\odot$ pc$^{-2}$. Of course, the mass of a galaxy is difficult to determine - only luminosity data is directly available. If the core of the galaxy were entirely composed of sun-type stars, the mean surface luminosity would equal 9600 $L_\odot$ pc$^{-2}$. However, the cores of most galaxies are very old, of the order of 10 billion years. The stars must therefore also be very old with a much lower light-to-mass ratio than the sun, which is only about 5 billion years old. Light from the centre also tends to be obscured by dust.

There is some evidence that the central luminosity of spiral galaxies is constant but at the level of 145 $L_\odot$ pc$^{-2}$ (Freeman,[6] but this has been described as a selection effect by Disney[7]). This is possibly consistent with the prediction if a population of older stars are assumed or there is a huge amount of non-luminous matter in the core of galaxies.

The nuclear bulge of many galaxies is not spherical but ellipsoidal or bar shaped. The mean mass surface density is independent of topology but the extremities of the bar are expected to form the transition between the bound and unbound states. The critical condition $GM/r^2 = 2a_o$ which then exists at the end of the bars can change with time. The acceleration $a_o$ decreases with time and $GM/r^2$ can vary through a variety of processes:

- A change in the value of G with time
- Mass loss from jet action at the galaxy core if the nucleus is active
- Equivalent mass loss through radiation

The first option is not significant - there is no evidence the gravitation constant varies, but undoubtedly mass can change. If the total mass loss per year is greater than about 10 solar masses, the gravitational acceleration at the ends of the bar falls below threshold with the result that material previously bound becomes free and begins to move outwards as the Hubble velocity kicks in.

The formation of spiral arms in disk galaxies is complex but we would expect the arms to originate from the ends of the bars through this process. For example, a near infra-red image of the core of M51 (the Whirlpool Galaxy, NGC 5194/5195) reproduced in Seeds, p234 [8], shows two spiral arm winding deeply into the core of the galaxy to the ends of a small bar. The outwards movement of material subsequent to 'ejection' is at low velocity and would be difficult to detect from redshift observations.

We can check the plausibility of this scenario with a rough calculation of the properties of the Milky Way galaxy (data taken from Seeds[8]): The radius of the nuclear bulge is about 3 kpc (9.2 x 10$^{19}$ m). The mass is estimated at 1.86 x 10$^{41}$ kg. The boundary acceleration is therefore 14.6 x 10$^{-10}$ m s$^{-2}$, very close to 2 times $a_o$. The Hubble radial velocity of the sun now (at a distance of 8.5 kpc) is about 0.6 km s$^{-1}$. Since this will remain constant in the simplest analysis (see Section 2), the time taken to move a distance 5.5 kpc to its current position is 8.8 billion years. We can say, roughly, that the sun is made from material that escaped the galactic nucleus when the Universe was 5 billion years old. This is of course a very crude estimate as the size of the nuclear bulge would have been considerably greater (because of greater mass) at the time of ejection.

This analysis points to anomalous behaviour associated with rotationally bound galaxy cores at radial distance of 1 – 4 kpc and may provide an physical basis for the effects MOND attempts to explain [2].

The flat rotation curves of disk galaxies can be explained by dark matter but requires unusual cooperation between dark and baryonic matter (the 'disk-halo conspiracy' [9]). The endpoint of the rotation curves is a function of luminosity only (the Tully-Fisher relation) and puts severe constraints on the nature of the dark matter halo[10]. Detailed application of the Hubble acceleration can explain the flat rotation curves in a fairly straightforward way:

Consider a spherical mass of constant density $\rho$. If the condition at the boundary of the sphere is

$$\frac{4\pi\rho Gr}{3} = \frac{2c}{T}$$

(12)

an unusual parasitic effect occurs where the core will continuously lose mass. As $T$ increases, $r$ decreases and the sphere steadily contracts whilst maintaining the equality of equation (12). The galaxy can be steadily and continuously depleted of central mass.

This will never occur with a large elliptical galaxy because the acceleration at the outer surface is greater than $2a_o$ and will normally always remain so (because $a_o$ is decreasing with time). If however the core becomes active and a large quantity of mass is ejected, the acceleration at the outer surface can fall to the critical level. At that stage mass will escape and spiral arms begin to develop. Once the process of disk formation begins, it is not easily terminated.

It should be noted that the size of gravitational acceleration at points within a stable galaxy can fall below $2a_o$, but this is not significant. It is not the magnitude of acceleration that defines the escape condition, but the binding energy. An acceleration of $2a_o$ is significant only because it corresponds to the energy limit *at the edge of the gravitation well* formed by the total bound mass.

It is possible to show that the rotation speed of the ejected material will be constant at any time by calculating the rotational history of mass escaping at an arbitrary time $T$. Assuming the mass was rotationally bound before escape, the initial rotational speed, $u$, is defined by the following expression:

$$\frac{4\pi\rho Gr}{3} = \frac{u^2}{r}$$

(13)

Rearranging,

$$u = \frac{2c}{T}\left(\frac{4\pi\rho G}{3}\right)^{-\frac{1}{2}}$$

(14)

The radial path can be modelled by assuming constant velocity $v_H$. Because angular momentum is conserved, the transverse velocity variation with time is then easily calculated. The initial angular momentum at escape point is (from equations (12) and (14))

$$L = mur = m\left(\frac{2c}{T}\right)^2 \left(\frac{4\pi\rho G}{3}\right)^{-\frac{3}{2}}$$

(15)

At current time $T_o$, the mass has moved to position $rT_o/T$. By equating the current and initial angular momenta, an expression that includes the current rotation speed, $u_o$, is derived:

$$u_o T_o = 2c \left(\frac{4\pi\rho G}{3}\right)^{-\frac{1}{2}}$$

(16)

The rotation speed is independent of radial position, although it declines steadily over time. In this analysis, information about central density variation is locked into the rotation curve as a deviation from the flat velocity curve.

However, real curves are not flat near the centre and it is unlikely that the density of a galaxy is constant. The predictive failure of equation (16) lies in the assumptions that have been made. A more realistic calculation must include the centripetal acceleration. It is found that the centripetal acceleration adds a small radial velocity, which is more significant when the distance from the centre is small. The small size of the additional outward velocity may seem surprising - one would expect the centripetal acceleration should quickly give rise to an overwhelming radial velocity. High-speed outward movement of this type would be apparent and simply does not occur.

In fact, the Hubble acceleration prevents the velocity growing through equation (4). As an outward peculiar velocity grows, an opposing Hubble acceleration builds up to balance the centripetal acceleration (this is similar to the engineering principle of 'proportional control' where the induced peculiar velocity would be labelled the 'error'). The additional outward velocity is much smaller than the Hubble velocity when the centripetal acceleration (minus the gravitational acceleration) falls significantly below $a_o$.

The procedure conserves energy and angular momentum. At distance r, let the centripetal force be balanced by the Hubble force

$$\frac{mu^2}{r} = \frac{mc}{T}\left(\frac{v_P}{v_H + v_P}\right)$$

(17)

The increase in Hubble Energy is

$$\frac{dE}{dt} = \frac{mcv_P}{T}$$

(18)

This is extracted from the rotational kinetic energy hence

$$mu\frac{du}{dt} = -\frac{mcv_P}{T}$$

(19)

The change in angular momentum is

$$\frac{dL}{dt} = mr\frac{du}{dt} + mu\frac{dr}{dt} = -\frac{mcv_P r}{uT} + mu(v_P + v_H)$$

(20)

Substituting equation (17), the RHS simplifies to 0. The Hubble force therefore conserves angular momentum. The process moves angular momentum outwards from the centre of the galaxy.

With the assumption that $v_H \gg v_P$, true with r large in the current epoch, equation (17) simplifies to $u^2 = v_P c$. A larger $u$ will result in a greater peculiar velocity and thus a greater reduction in $u$. A detailed analysis shows that the contribution from the centripetal acceleration to the rate of change of $u$ under these circumstances is actually $-u^3/rc$, with the result that slightly rising curves are also possible.

The process may explain the rotation curve but not the visible arms of a galaxy. Possibly, the centripetal acceleration can drive a much faster acoustic wave.

Flat rotation curves in disk galaxies are therefore not unusual – they may be inevitable. This is a simpler explanation than invoking dark matter or altering the gravitational force equation.

Looking again at equation (9) in light of energy conservation; where does the Hubble energy go when a mass makes the transition from the free to bound state? It is proposed that the energy becomes associated with the gravitational system and has a cosmological significance. If a mass at the threshold acceleration is captured at time $T_1$ then escapes at time $T_2$ (captured and emerging in the same place), there is a nett loss of energy of $mc^2 (1/T_1 - 1/T_2)$. During the time the mass was bound, the universe has expanded and the gravitational energy of the universe has reduced. It is possible that on a global level the effects balance.

**5. Energy Loss with Redshift**

A major source of concern with the big bang model is the apparent violation of the principle of energy conservation by the cosmological expansion. The received energy is less than that emitted, in the case of the cosmic background radiation, by a factor of over one thousand.

Consider the case of photon emission to a distant absorber moving away from the emitter with a Hubble velocity $v$. In the rest frame of the emitter, the emitted energy is $E_{em}$. It is known both from relativity and experimentally that the absorbed energy $E_{abs}$ (in the rest frame of the absorber) is

$$E_{abs} = \frac{E_{em}}{(1+z)}$$

(21)

Because the absorber is moving radially away at velocity $v$, the apparent energy gain at the absorber from the viewpoint of the emitter, $E_{abs}^{em}$, is Lorenz boosted

$$E_{abs}^{em} = \gamma \frac{E_{em}}{(1+z)}$$

(22)

This is still less than the emitted energy hence there is an apparent unexplained energy loss associated with the photon transfer event when considered from the emission frame. The energy associated with the recoil velocity has been ignored as this can be made arbitrarily small by a suitable choice of emitter and absorber masses.

The discrepancy is surprising. It is true of Special Relativity that observers in different frames can measure different energies for the same event (because of the rotation of Energy-Momentum vector) but all will observe energy and momentum conservation in any energy transfer process. This problem with the cosmological redshift is accepted in cosmology because General Relativity does not absolutely admit energy conservation unless some sort of curvature energy is ascribed to the metric (this has yet to be quantitatively demonstrated). This is paradoxical because the Friedmann equation that describes the universe is derived on the basis of energy conservation (the kinetic energy of the Hubble flow loses energy as it works against gravity). It should be noted that the gravitational redshift does conserve energy – the redshift is associated with real energy loss as the photon escapes the gravitational energy well. Historically, a refusal to accept energy non-conservation led to the discovery of the neutrino.

The introduction of the Hubble energy resolves the problem and restores energy conservation. If we assume Hubble energy as defined in equation (2), the Hubble energy of the absorber will increase because of the equivalent mass increase associated with the energy transfer. The total energy change at the absorber from the emitter frame is now correctly expressed as

$$E_{abs}{}^{em} = \gamma \frac{E_{em}}{(1+z)}(1+\frac{v}{c})$$

(23)

The RHS of the equation simplifies to $E_{em}$ - energy is conserved.

Now consider the situation from the viewpoint of the absorber. This is the viewpoint when observing distant objects from the Earth. From the previous argument we know the received energy is reduced and is consistent with that actually measured, but from the absorber frame, the situation should be very different. The distant photon will have additional energy because the Hubble energy of the source recession velocity is recovered. Thus the received energy should be greater by a factor (1 + v/c). Why is it not? The point is that the situation is not symmetrical. The symmetry is broken by both the expansion of the universe and the time arrow – photons travel in the direction of positive time only. Thus, the absorber is moving away from the emitter at velocity $v$, but the emitter is not moving away from the observer at velocity $v$ (even though the relative velocity is certainly $v$). Once a photon has left the emitter, the subsequent movement of the emitter is irrelevant. Not so with the absorber – the velocity history affects propagation time. The situation is clearly not symmetrical.

There is a preferred frame, but in cosmology this is acceptable; for example the cosmic background radiation defines a zero velocity reference frame against which the absolute motion of the Earth through space is measured.

In this paper, the Hubble energy has been described from the frame of the emitter. We could equally well have described it from the frame of the absorber - the energy is the same magnitude, but negative.

## 6. Discussion

The Hubble Energy has been introduced in a simple way and has been shown to offer qualitative explanations to some outstanding cosmological problems.

It should be noted that the Hubble Energy, if it exists, should affect the development of the Universe. The universe is modelled with the Friedmann Equation. A simple way to derive the equation is by imagining a sphere expanding with time and applying energy conservation at the extreme edge [11]. The resulting equation is normally modified to include a cosmological term with characteristic constant lambda (Λ):

$$\dot{R}^2 = \frac{8\pi G \rho R^2}{3} + \frac{\Lambda R^2}{3} - \kappa c^2$$

(24)

$R$ is the radius of the universe, $\rho$ is the mean density, and $\kappa$ is the curvature. The current belief is that the universe is flat, hence $\kappa = 0$.

By attributing Hubble energy to the edge of the sphere, the Friedmann equation modifies to

$$\dot{R}^2 = \frac{8\pi G \rho R^2}{3} - 2\dot{R}c + \frac{\Lambda R^2}{3} - \kappa c^2$$

(25)

A new parameter, Γ, with the same form and dimension as Λ can be defined:

$$\Gamma = -6\frac{\dot{R}c}{R^2}$$

(26)

# Hubble Energy

The modified Friedmann equation can now be rewritten in terms of Γ:

$$\dot{R}^2 = \frac{8\pi G \rho R^2}{3} + \frac{(\Lambda + \Gamma)R^2}{3} - \kappa c^2$$

(27)

Fitting plausible values to equation (26), we can estimate that Γ has a value of about $-3 \times 10^{-35}$ s$^{-2}$. It may be tempting to identify gamma with the existing cosmological constant but analysis of the WMAP cosmic background data [12] suggests a positive value of lambda (although of similar magnitude). $\Lambda/3H^2$ was found to be 0.73 and the equivalent value of Λ is $+1.1 \times 10^{-35}$ s$^{-2}$.

As a consequence, the Hubble energy would seem not to be related to the postulated dark energy. However, the Hubble energy was added to the Friedmann Equation from the viewpoint of the emitter. We actually view the universe as the absorber, when the sign of the Hubble energy changes. A more precise analysis is required to see if there is any connection between the Hubble and dark energies.

There is a range of conceptual issues associated with the Hubble energy but these are not considered in this introductory paper.